\documentclass[aip,pop,preprint,superscriptaddress]{revtex4-1}
\usepackage{lineno,hyperref}
\modulolinenumbers[5]

\usepackage{graphicx}
\usepackage{amsmath,amssymb}
\usepackage{color}
\begin{document}

%%%%%%%%%%%%%%%%%%%%%%%%%%%%%%%%%%%%%%%%%%%%%%%%%%%%%%%%%%%%%%%%%%%%%%%%%%%%%%%%%%%%%%%%%%%%%%%%%%%%%%%%%%%%%%%%%%%%%%%%%%%%%%%%%%%%%%%%%%%%%%
\title{Role of magnetic field curvature in magnetohydrodynamic turbulence}

\author{Yan Yang}
\affiliation{Southern University of Science and Technology, Shenzhen, Guangdong 518055, China}
\affiliation{University of Science and Technology of China, Hefei, Anhui 230026, China}
\author{Minping Wan}
\email{wanmp@sustech.edu.cn}
\affiliation{Southern University of Science and Technology, Shenzhen, Guangdong 518055, China}
\author{William H. Matthaeus}
\affiliation{University of Delaware, Newark, DE 19716, USA}
\author{Yipeng Shi}
\affiliation{Peking University, Beijing 100871, China}
\author{Tulasi N. Parashar}
\affiliation{University of Delaware, Newark, DE 19716, USA}
\author{Quanming Lu}
\affiliation{University of Science and Technology of China, Hefei, Anhui 230026, China}
\author{Shiyi Chen}
\affiliation{Southern University of Science and Technology, Shenzhen, Guangdong 518055, China}

\date{\today}

\begin{abstract}
Magnetic field are transported and tangled by turbulence, even as they
lose identity due to nonideal or resistive effects. On balance field
lines undergo stretch-twist-fold processes.
The curvature field, a scalar that measures the tangling of the
magnetic field lines, is studied in detail here,
in the context of magnetohydrodynamic turbulence.
A central finding is that
the magnitudes of the curvature and the magnetic field are anti-correlated.
High curvature co-locates with low magnetic field, which
gives rise to power-law tails of the probability density function
of the curvature field. The curvature drift term that converts magnetic
energy into flow and thermal energy, largely depends on
the curvature field behavior, a relationship that
helps to explain particle acceleration due to curvature drift.
This adds as well to evidence that turbulent effects most likely play an
essential role in particle energization
since turbulence drives stronger tangled field configurations,
and therefore curvature.
\end{abstract}

%\begin{keyword}
%magnetohydrodynamic turbulence, magnetic reconnection, curvature drift, acceleration of particles
%\end{keyword}
\maketitle

\section{Introduction}
A divergence-free vector field, such as the magnetic field, can be conveniently
visualized in terms of field lines, which are tangent to the field everywhere.
In many astrophysical and space plasmas, magnetic field lines play an
essential role, in general for describing topology and connectivity
at a single instant of time, and even for developing  theoretical
descriptions of dynamical processes such as magnetic reconnection \cite{Parker-cmf}.
Magnetic field lines are also widely used in
describing mechanisms for particle acceleration and for
the transport of suprathermal and
energetic particles.
Ambiguities in defining field lines, especially as
a function of time, are well known \cite{Newcomb58}.
% W. A.Newcomb. Annals Physics,3, 348 (1958)
In turbulence, magnetic field lines are in geneal
not well-ordered, but rather
exhibit complex structures and wander randomly in space
\cite{Jokipii1966ApJ,jokipii1969stochastic,matthaeus1995spatial}.
Moving into the realm of magnetic reconnection, the field lines
can ``{\it disconnect}''
and ``{\it reconnect}''.
Given these ambiguities inherent in the magnetic field line
formalism,
here we avoid committing to a focus on the trajectory
of the magnetic field integral curves,
and instead prefer to consider
an intrinsic geometric parameter: curvature,
that completely determines a curve in 2D space.

The curvature $\kappa$, measuring how rapidly
a curve changes direction in space, is defined as
\begin{equation}
\kappa=\left\| \boldsymbol{b} \cdot \nabla \boldsymbol{b} \right\|.
\end{equation}
where $\boldsymbol{b}=\boldsymbol{B}/\left\|\boldsymbol{B}\right\|$.
The curvature of magnetic field lines is related directly
to the curvature drift of the motion
of charged particles, which is invoked in certain
particle acceleration mechanisms, e.g., first-order Fermi mechanism in magnetic reconnection
\cite{northrop1961guiding,drake2006electron}.
Indeed, rapid advances in computations and observations
have improved understanding of several features of magnetic reconnection
\cite{hoshino2001suprathermal,egedal2012large,drake2006electron,oka2010electron,pritchett2006relativistic}
that might contribute to particle energization.
For example, the curvature drift mechanism (related to the first-order Fermi mechanism) has
been identified as
the dominant source of electron heating with a weak guide field
\cite{dahlin2014mechanisms,dahlin2017role,li2017particle,li2015nonthermally,guo2014formation,lu2018formation,wang2016mechanisms,wang2017electron}.
One might reason in a qualitative sense: we think of bent field lines as elastic bands
under tension, exerting a force ($\sim \kappa B^2$) on the fluid.
This force will drive flows as the lines straighten out. It is natural
then to inquire the extent to which the curvature drift
acceleration and the curvature field
are spatially and quantitatively correlated.

Although there is hardly any doubt that the curvature, as a
essential feature of the geometrical behavior of magnetic field lines,
is a key ingredient in certain particle acceleration mechanisms,
the detailed properties of the magnetic field curvature is neither
well known nor well understood for general configuration.
The study presented in this letter
addresses this problem by employing
numerical simulations of magnetohydrodynamic (MHD) turbulence,
with the goal of providing better physical
interpretation of curvature drift acceleration.
We find that intense curvature and small magnetic
field are preferentially colocated.
Intense curvature tends to be
linearly anti-correlated with the square of
magnetic magnitude.
Therefore, magnetic energy release via the curvature drift term
is in strong association with the curvature field.
This result provides
new insights into the
literature on particle energization during magnetic reconnection,
and also favors the influence of turbulent magnetic effects on heating.

\section{Method}
We consider two-dimensional incompressible MHD turbulence.
The dynamical equations read
\begin{eqnarray}
{\frac{\partial \omega}{\partial t}} + \left(\boldsymbol{v} \cdot \nabla\right) \omega &=& \left(\boldsymbol{B} \cdot \nabla\right) j + \nu \nabla^2 \omega, \\
{\frac{\partial a}{\partial t}} + \left(\boldsymbol{v} \cdot \nabla\right) a &=& \eta \nabla^2 a,
\end{eqnarray}
where $\boldsymbol{v}$ is the velocity field,
$\omega=(\nabla \times \boldsymbol{v}) \cdot \hat{\boldsymbol{z}}$ is the vorticity,
$a$ is the magnetic potential, $\boldsymbol{B}=\nabla a \times \hat{\boldsymbol{z}}$ is the magnetic field,
and $j=(\nabla \times \boldsymbol{B}) \cdot \hat{\boldsymbol{z}}=-\nabla^2 a$ is the electric current density.
For simplicity, the viscosity $\nu$ and resistivity $\eta$ are set to equal values.
The numerical simulation is done with a Fourier spectral method \cite{wan2013generation} in a doubly
periodic $(2\pi)^2$ cartesian domain with
a $8192^2$ resolution. The fields are initialized at modes $5\le |\boldsymbol{k}| \le 20$
with random phases and fluctuation amplitudes, whose spectra are proportional to
$1/\left[1+(k/k_0)^{8/3}\right]$ with $k_0=10$. The total kinetic and magnetic energy are
each initially equal to $0.5$. We fix $\nu=\eta=5 \times 10^{-5}$,
which corresponds to high
Reynolds number.
We carry out our analysis on snapshots
near the time of maximum mean square electric current density.

\section{Anti-correlation between curvature and magnetic field}
\label{sec:2D-k-b}
\begin{figure}%[!htpb]
\centering
\includegraphics[width=0.5\textwidth]{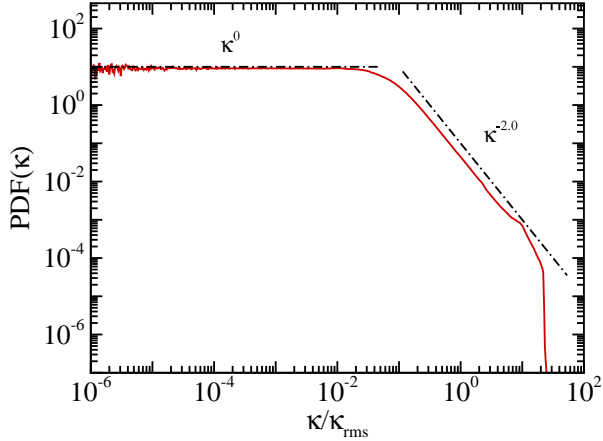}
\caption{PDF of the magnetic field curvature $\kappa$ normalized to the root
mean square value $\kappa_{\rm rms}$. The PDF has a $\kappa^0$ low-curvature regime
and a $\kappa^{-2}$
high-curvature tail.}
\label{Fig.PDF-k}
\end{figure}

Curvature is a well-studied geometric characteristic
of particle trajectories, its statistical
properties having been reported in numerous
hydrodynamic turbulence studies
\cite{braun2006geometry,XuHaitao07,ouellette2007curvature,ouellette2008dynamic,scagliarini2011geometric,
kadoch2011lagrangian,yang2013acceleration}.
Particle trajectories are manifestly different from magnetic field lines,
but similar analysis methods are useful to study both cases.
As a first step we compute the
probability distribution function (PDF) of field line curvature in the above-described MHD simulation.  The results,
shown in Fig. \ref{Fig.PDF-k}, reveal
properties that are surprisingly reminiscent of
the hydrodynamic particle trajectory case.
The distribution is broad and
exhibits two clear power-law regimes --
a low-curvature plateau scaling as $\sim \kappa^0$, while,
the distribution of large curvatures scales as $\sim \kappa^{-2}$.

The physical origins
of this feature become more apparent when we rewrite the Lorentz force
in terms of $\boldsymbol{B}$,
$\boldsymbol{j} \times \boldsymbol{B}=\boldsymbol{B} \cdot \nabla \boldsymbol{B} -\nabla \left(\boldsymbol{B}^2/2\right)$,
where the second term acts in the same way as the pressure force
and the first term is equivalent to $\nabla \cdot \left(\boldsymbol{B}\boldsymbol{B}\right)$,
which can be interpreted as the effect of the surface stress $B_iB_j$ on fluids.
Then we rewrite the force
$\boldsymbol{B} \cdot \nabla \boldsymbol{B}$ in terms of
curvilinear coordinates attached to a field line,
\begin{equation}
\boldsymbol{B} \cdot \nabla \boldsymbol{B}=B{\dfrac{\partial B}{\partial s}} \boldsymbol{t} -{\dfrac{B^2}{R}} \boldsymbol{n}.
\end{equation}
Here $B=\left\|\boldsymbol{B}\right\|$, $s$ is a coordinate along the field line,
$\boldsymbol{t}=\boldsymbol{b}$ and $\boldsymbol{n}$ are unit vectors
in the tangential and normal directions, respectively,
and $R=1/\kappa$ is the local field line radius.
It follows that the curvature can be expressed as
\begin{equation}
\kappa=\frac{\left\|\boldsymbol{b} \times \left(\boldsymbol{B} \cdot \nabla \boldsymbol{B}\right) \right\|}{B^2}=\frac{f_n}{B^2}.
\label{Eq.k}
\end{equation}

\begin{figure*}%[!htpb]
\centering
\centering
\includegraphics[width=0.45\textwidth]{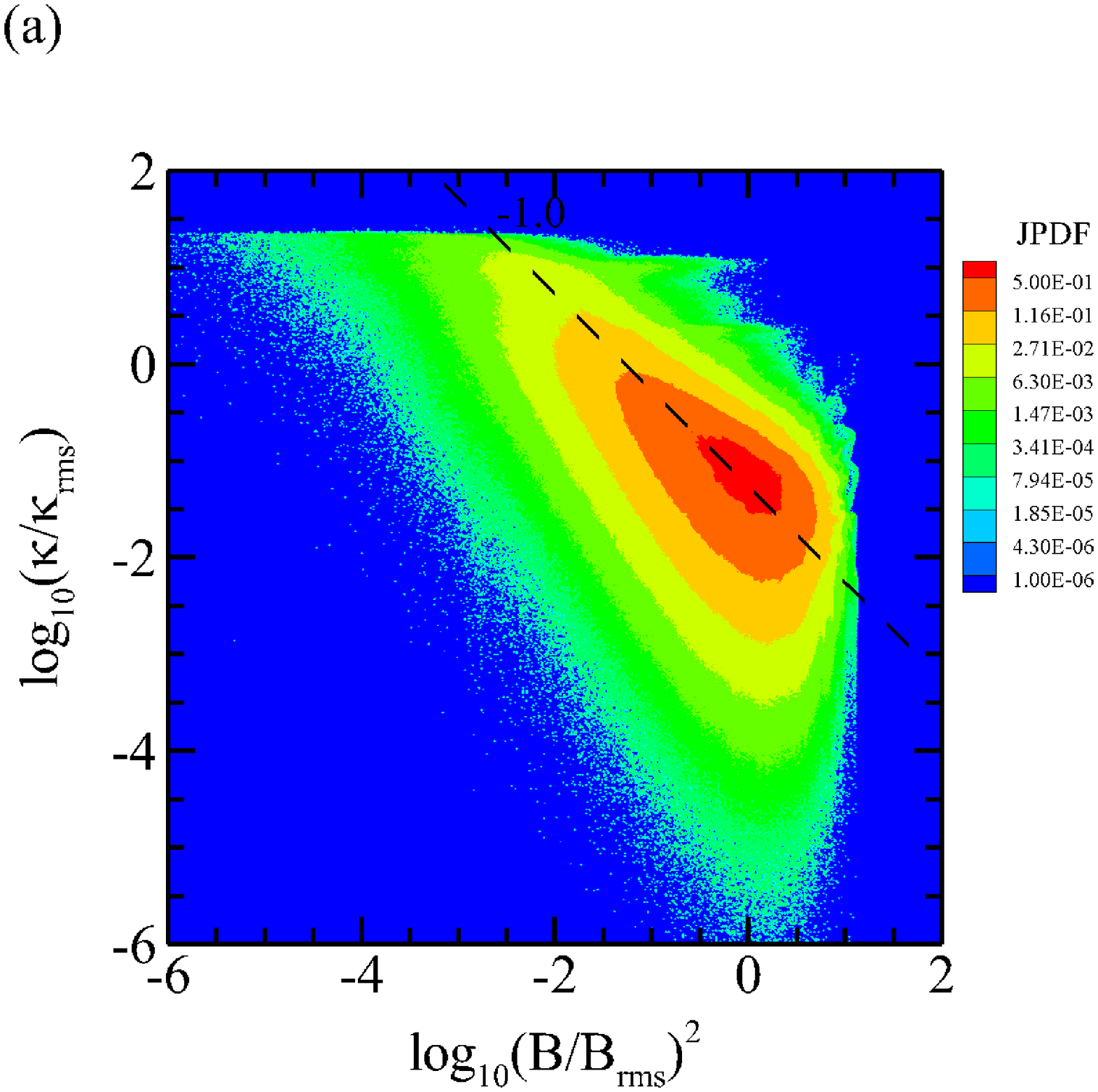}
\includegraphics[width=0.45\textwidth]{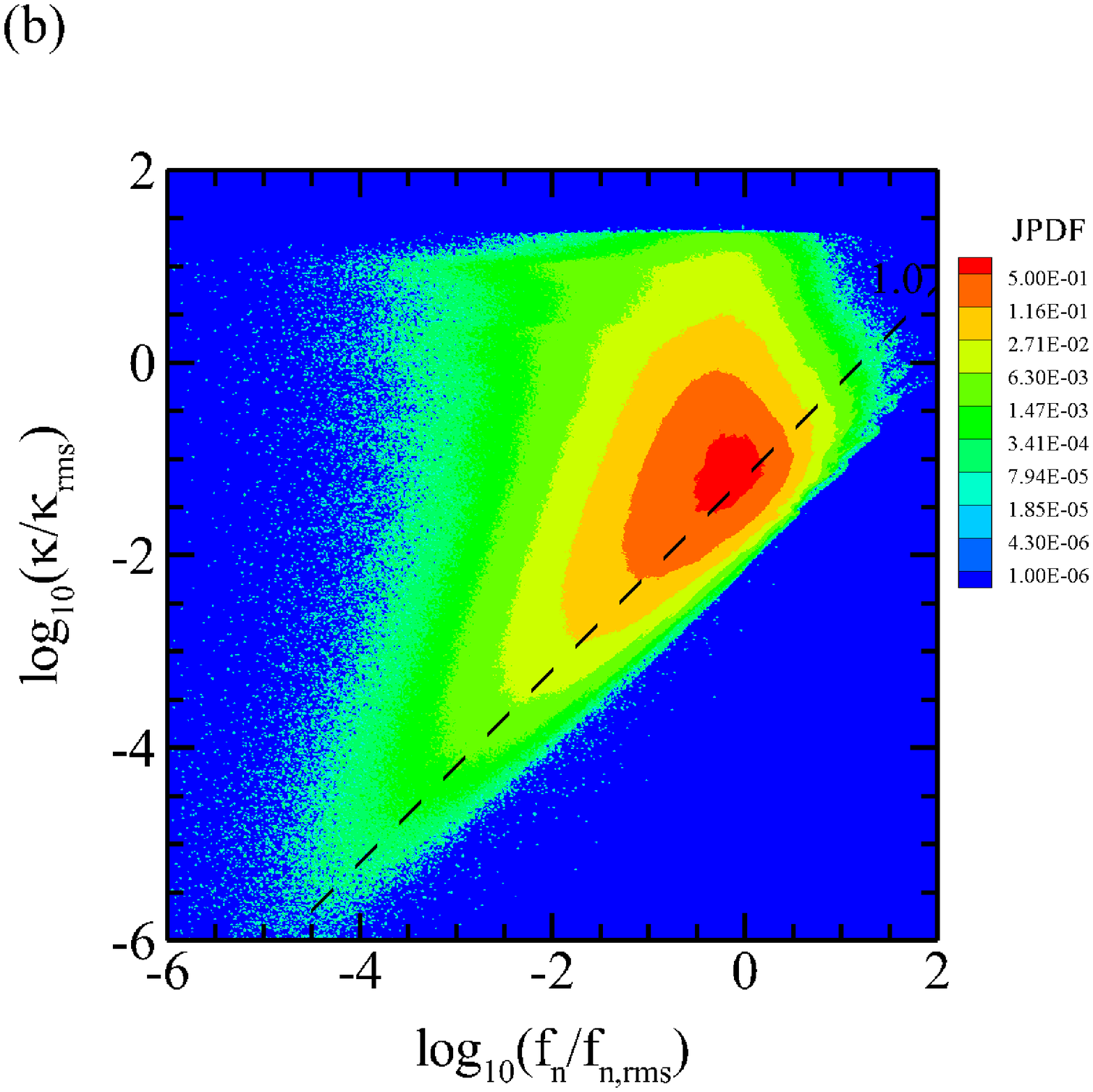}
\caption{Joint PDFs of the curvature $\kappa$ and
(a) the square of magnetic magnitude $B^2$ and (b) the force magnitude $f_n$ acting normal to
field lines. All quantities are normalized to their respective root mean square values.
There are apparent associations between high curvature and low magnetic field and between low curvature and low normal force.}
\label{Fig.JPDF-k-b}
\end{figure*}

\begin{figure*}%[!htpb]
\centering
\centering
\includegraphics[width=0.45\textwidth]{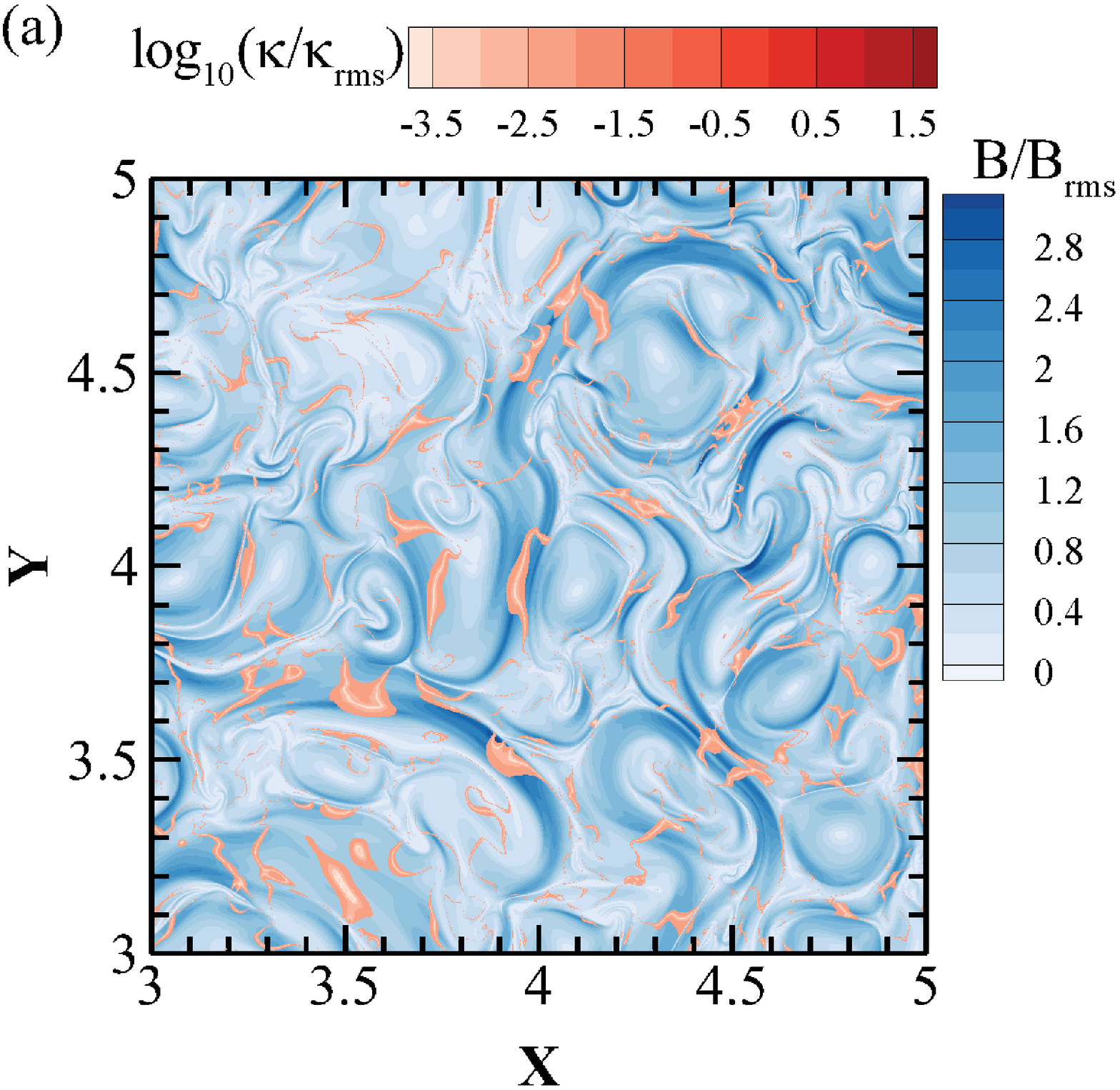}
\includegraphics[width=0.45\textwidth]{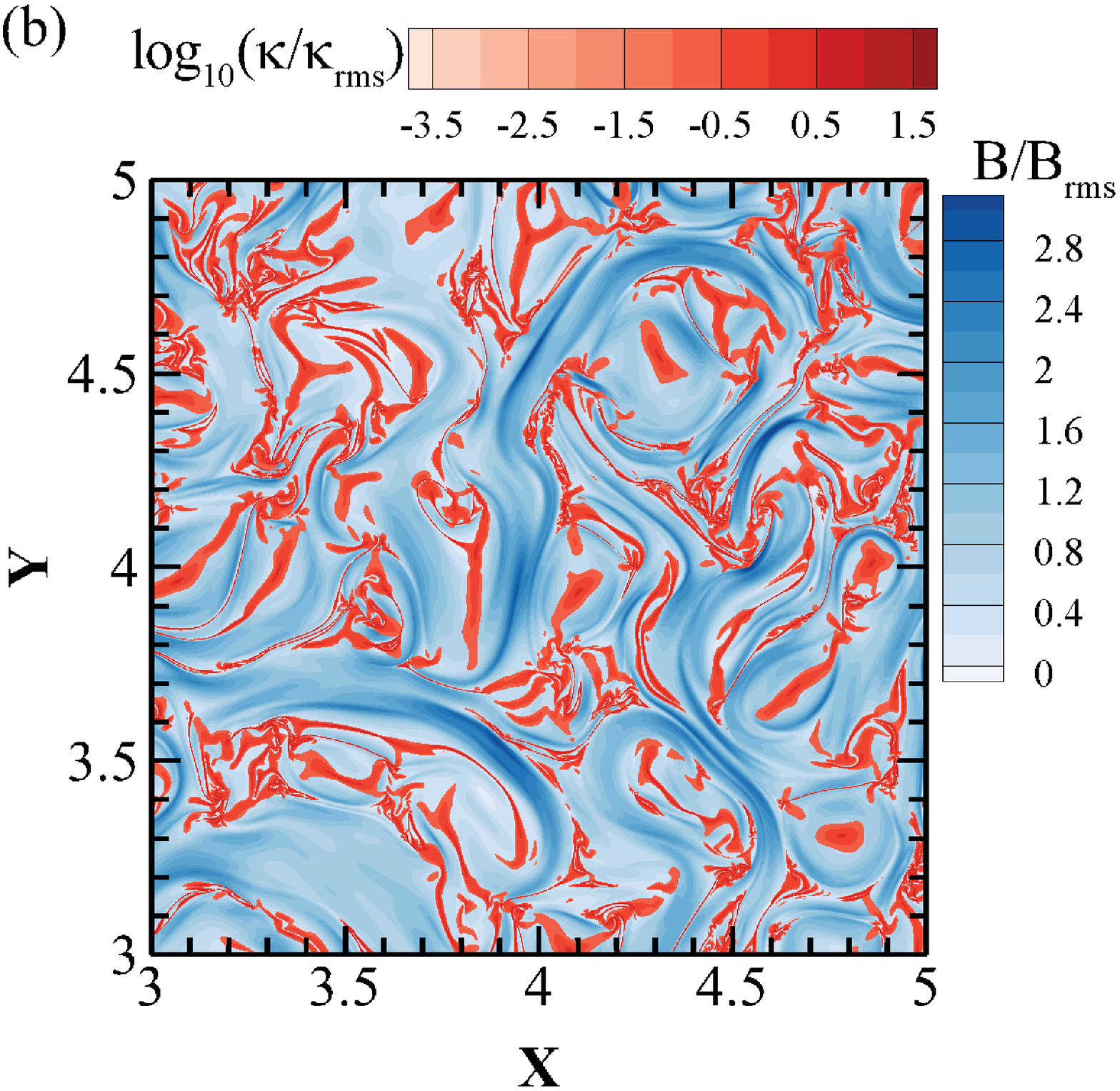}
\caption{Contour maps of curvature
superposed on color map of magnetic magnitude
in a subregion of the whole domain.
(a) Only $\kappa/\kappa_{\rm rms}<0.01$ values are pictured;
(b) Only $\kappa/\kappa_{\rm rms}>0.2$ values are pictured.
High curvature populates in regions of low magnetic field.}
\label{fig:xy-b-k}
\end{figure*}

According to Eq. \ref{Eq.k}, one might expect that
large normal force $f_n$ and small magnetic field both
correspond to high curvature.
However, from the joint PDFs in Fig. \ref{Fig.JPDF-k-b},
one can see that high curvature is not
strongly correlated with large normal force but instead
is well associated with small magnetic field magnitude,
while their effects are opposite for low curvature.
By locating the small and large values
of the curvature field,
values $\kappa/\kappa_{\rm rms}<0.01$ and
$\kappa/\kappa_{\rm rms}>0.2$ are plotted on the top of the color map
of magnetic magnitude in Fig. \ref{fig:xy-b-k},
which correspond to the $\kappa^0$ low-curvature and
$\kappa^{-2}$ high-curvature tails shown in Fig. \ref{Fig.PDF-k}, respectively.
No qualitative associations between low curvature and
large magnetic magnitude have been observed in Fig. \ref{fig:xy-b-k}(a),
while we can readily identify
the concentration of high curvature in regions
of low magnetic magnitude in Fig. \ref{fig:xy-b-k}(b).
These regions of low magnetic field strength
are organized into twisted lines and isolated points.
The solitary points are visually found in the vicinity of magnetic island cores,
where the direction of magnetic field
changes significantly, leading to high curvature.
The high-curvature lines are preferentially in the form
of sheet-like structures around the rims of islands,
reminiscent of the well-studied configuration
of sheets of electric current density.
It is likely, of course,
that this association is also related
to potential sites of
magnetic reconnection \cite{Parker-cmf}.

Based on the above results,
we expect that
the large-curvature (i.e. $\kappa \rightarrow \infty $) and small-curvature (i.e. $\kappa \rightarrow 0$)
regimes in Fig. \ref{Fig.PDF-k}
should be
determined by the scaling behavior of $1/B^2$ as $B \rightarrow 0$ and
$f_n$ as $f_n \rightarrow 0$, respectively.
To make these connections,
one may begin by recalling
that the $x$ and $y$ components of magnetic fluctuations
are independent in isotropic MHD turbulence,
and have quasi-Gaussian distributions.
The square of magnetic magnitude $B^2=B_x^2+B_y^2$ therefore follows
a chi-squared distribution with 2 degrees of freedom,
deriving the $\kappa^{-2}$ high-curvature tail as $B \rightarrow 0$
by using Taylor expansion.
Continuing, suppose we
assume that the force $f_n$ at low values
is a quasi-Gaussian random variable,
recognizing that
it could deviate from Gaussian distribution at high values due to intermittency.
Then the $\kappa^0$ low-curvature tail is recovered as $f_n \rightarrow 0$.

\section{Curvature drift acceleration}
The discovery of the anti-correlation between curvature and magnetic fields
is particularly suggestive of
some generalizable physical process, such as curvature drift acceleration.
From the Faraday's law, one readily finds
the equation governing magnetic energy
$E^m=B^2/2$,
\begin{equation}
\frac{\partial E^m}{\partial t} + \nabla \cdot \left(\boldsymbol{E} \times \boldsymbol{B}\right)=-\boldsymbol{E} \cdot \boldsymbol{j},
\label{Eq.em}
\end{equation}
wherein $\boldsymbol{E} \cdot \boldsymbol{j}=\boldsymbol{E} \cdot \boldsymbol{j}_{\parallel}+\boldsymbol{E} \cdot \boldsymbol{j}_{\perp}$,
and $(\cdots)_\parallel$ and $(\cdots)_\perp$ are quantities parallel and perpendicular with respect to the local magnetic field direction.
The perpendicular one can be further decomposed as
$\boldsymbol{E} \cdot \boldsymbol{j}_{\perp}= \boldsymbol{E} \cdot \left[\dfrac{\boldsymbol{B} \times \left(\boldsymbol{B} \cdot \nabla\right) \boldsymbol{B}}{B^2}\right] - \nabla\left(\dfrac{B^2}{2}\right) \cdot {\dfrac{\boldsymbol{E} \times \boldsymbol{B}}{B^2}}$,
where the second term on the right-hand side
can be combined with the second term on the left-hand side in Eq. \ref{Eq.em}.
The remaining term due to curvature drift is
\begin{equation}
\boldsymbol{E} \cdot \boldsymbol{j}_c = \boldsymbol{E} \cdot \left[\frac{\boldsymbol{B} \times \left(\boldsymbol{B} \cdot \nabla\right) \boldsymbol{B}}{B^2}\right] \sim \kappa B E_{\perp},
\label{Eq.curvature-drift}
\end{equation}
where $\boldsymbol{j}_c$ is the electric current density due to curvature drift.

\begin{figure*}%[!htpb]
\centering
\includegraphics[width=0.45\textwidth]{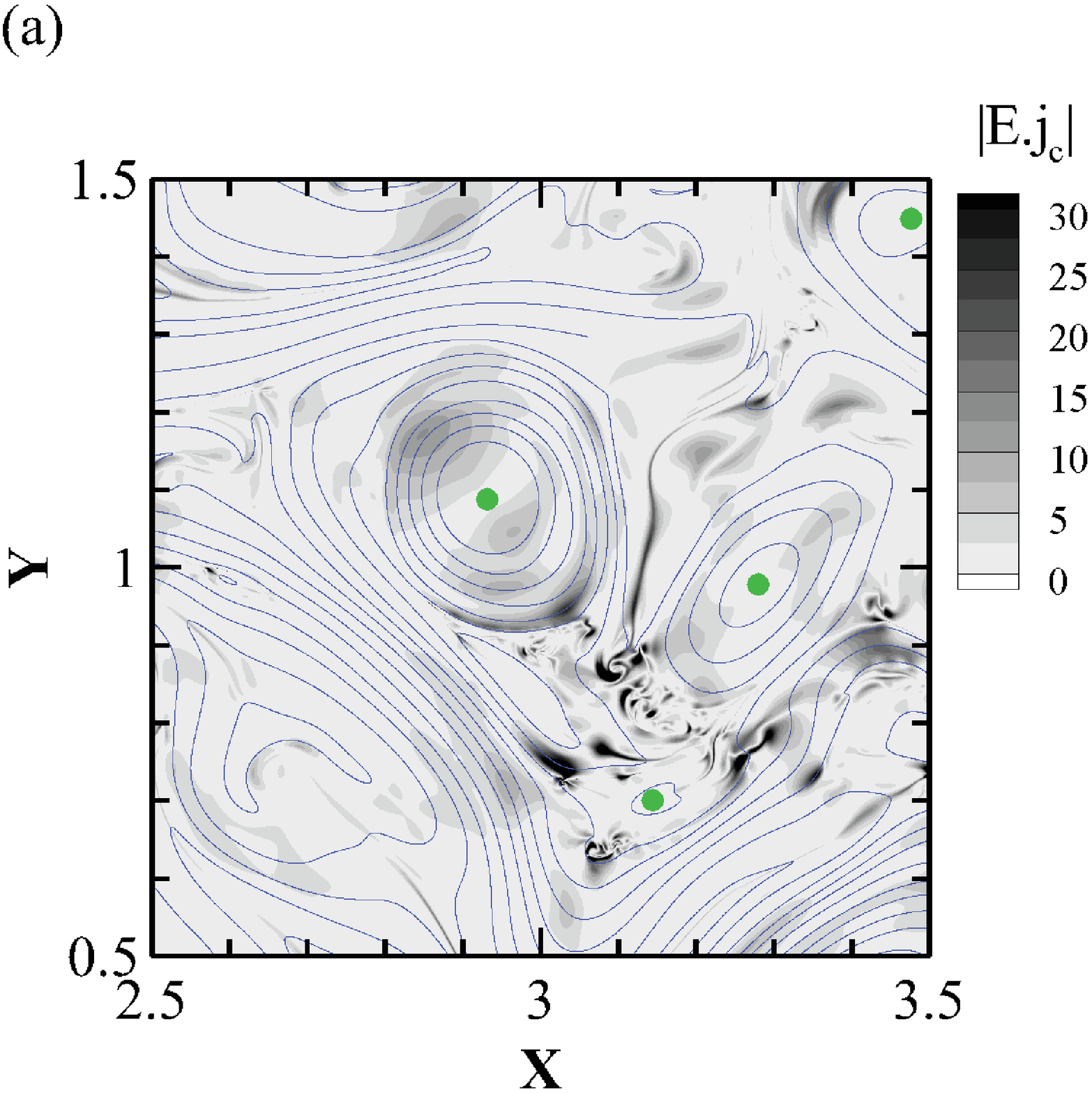}
\includegraphics[width=0.45\textwidth]{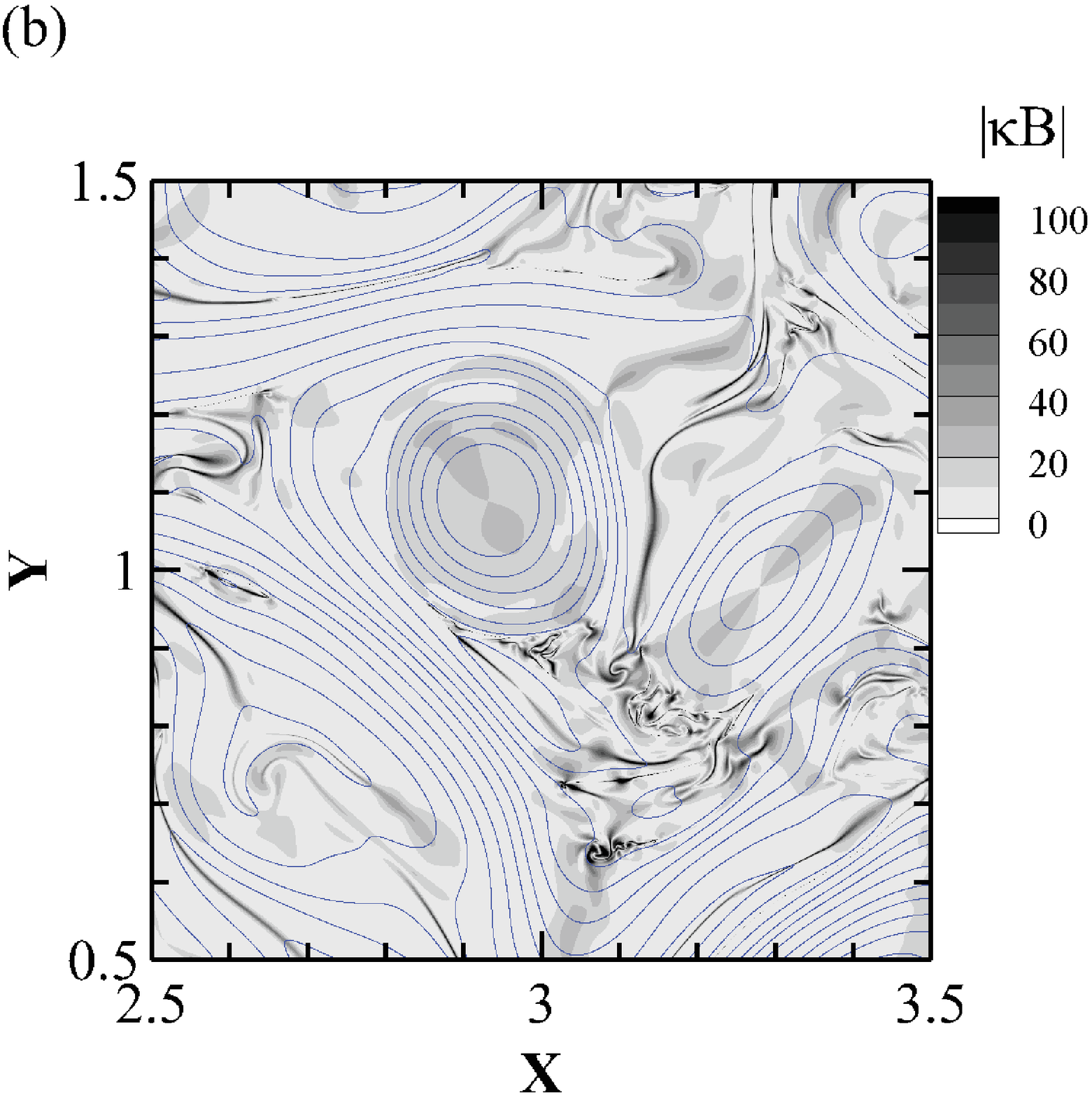}
\includegraphics[width=0.45\textwidth]{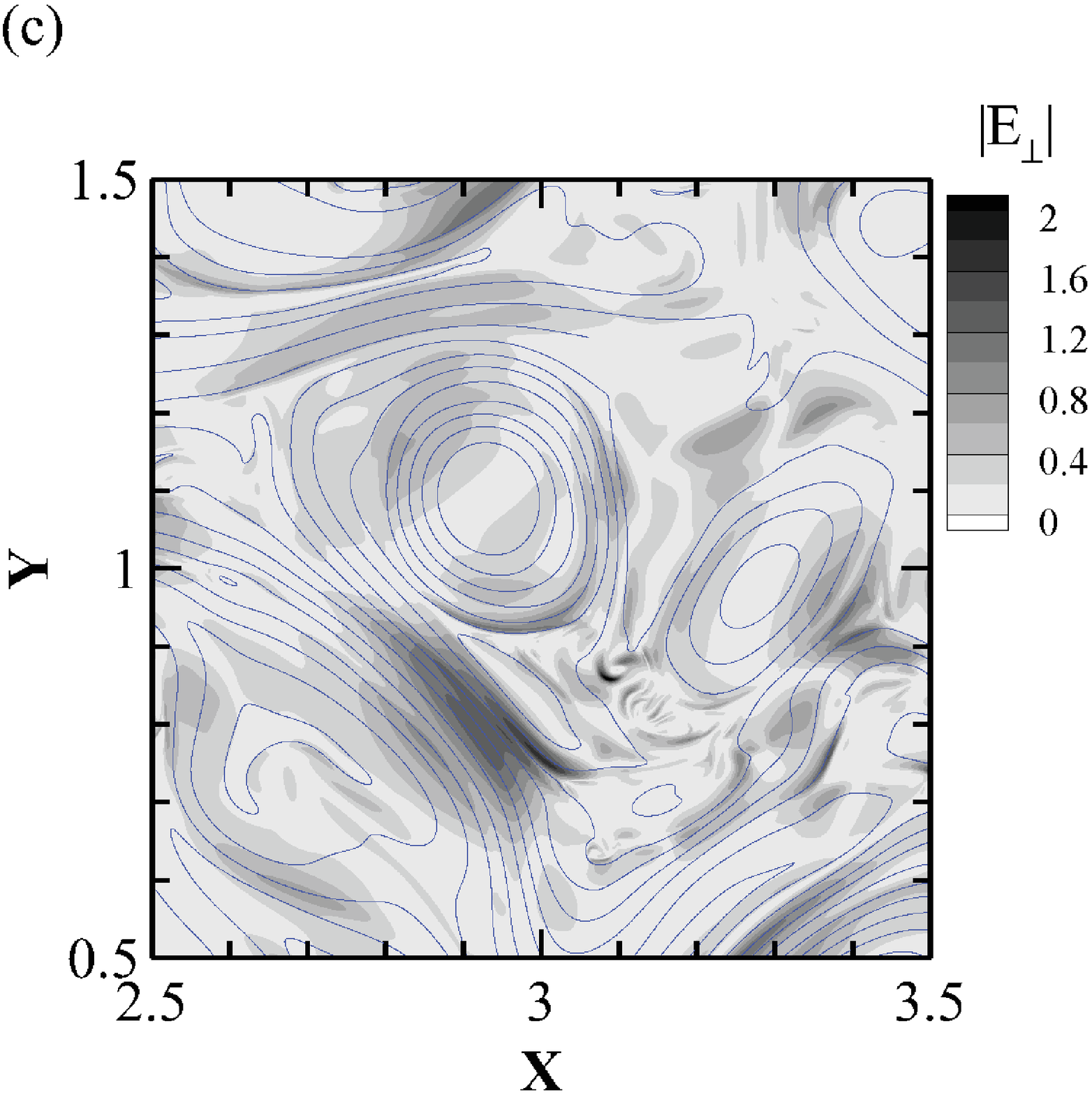}
\includegraphics[width=0.45\textwidth]{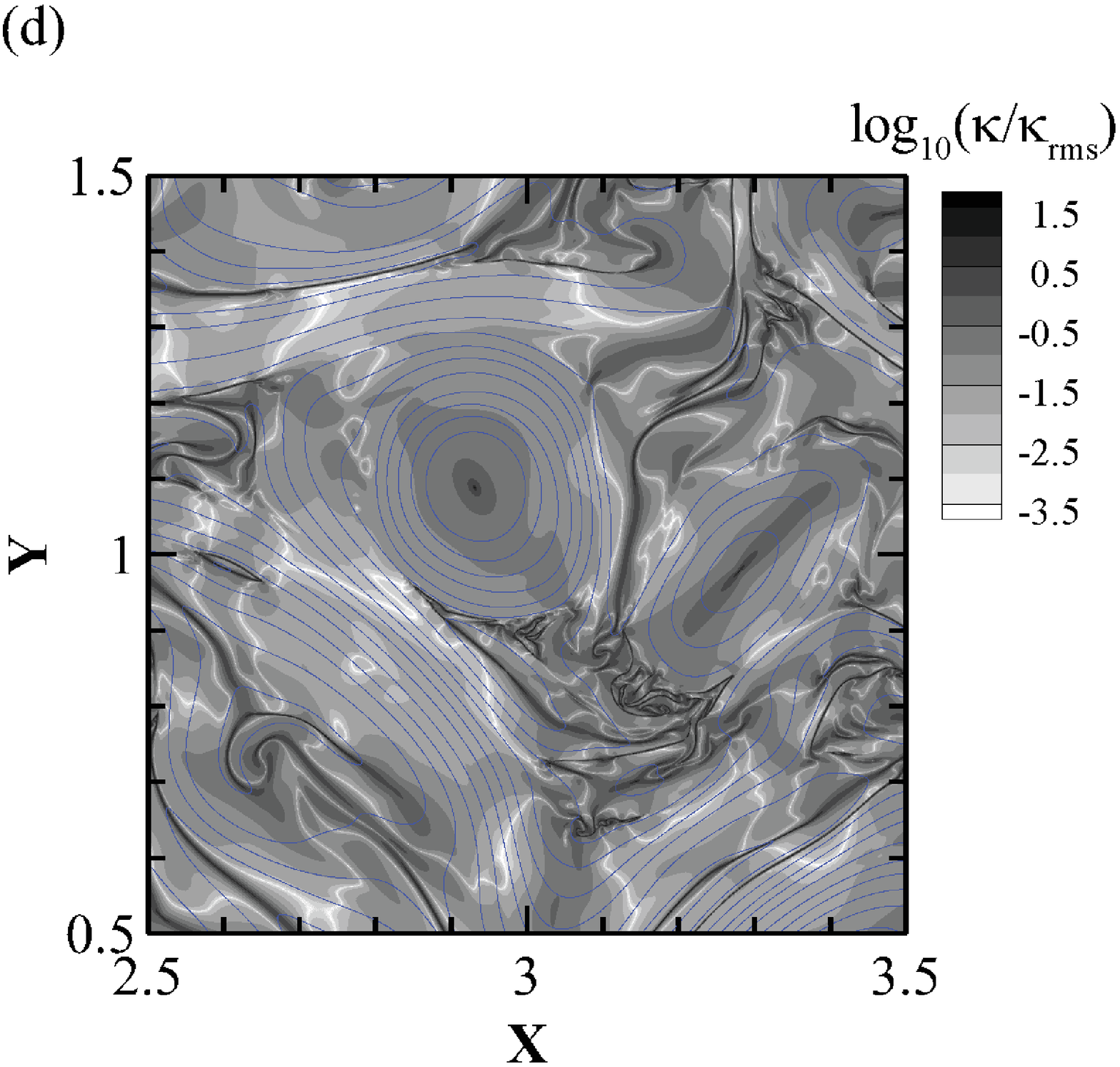}
\caption{Contour maps of (a) the curvature drift acceleration term $|\boldsymbol{E} \cdot \boldsymbol{j}_c|$,
(b) the curvature-related component $|\kappa B|$,
(c) the perpendicular electric field $|E_{\perp}|$, and
(d) the logarithm of the curvature $\kappa$
in a subregion of the whole domain with the blue contour lines showing the
magnetic field lines.}
\label{xy-em-k}
\end{figure*}

The spatial distributions of the curvature drift acceleration term
$\boldsymbol{E} \cdot \boldsymbol{j}_c$ (see Fig. \ref{xy-em-k}(a)) and
the curvature-related component $\kappa B$ (see Fig. \ref{xy-em-k}(b))
behave quite similarly over the whole domain, apart from regions
in the proximity of some special points like magnetic island cores
marked as green circles.
Instead, the curvature drift acceleration term
$\boldsymbol{E} \cdot \boldsymbol{j}_c$ and the perpendicular electric field $E_{\perp}$
(see Fig. \ref{xy-em-k}(c)) exhibit similarity near these special points.
We apply the method \cite{servidio2010statistics}, examining the topography of magnetic potential,
to find O-points (magnetic island cores) and X-points (magnetic reconnection sites) in 2D.
The electric field from Ohm's law
$\boldsymbol{E}=-\boldsymbol{v}\times \boldsymbol{B} + \eta \boldsymbol{j}$ is dominated
globally by the term $\boldsymbol{v}\times \boldsymbol{B}$,\cite{servidio2010statistics}
and which is
$\boldsymbol{E}= \eta \boldsymbol{j}$ at magnetic island cores and reconnection sites
since the magnetic field vanishes at these positions.
Therefore the perpendicular electric field near these special points is very small.

\begin{figure*}%[!htpb]
\centering
\centering
\includegraphics[width=0.45\textwidth]{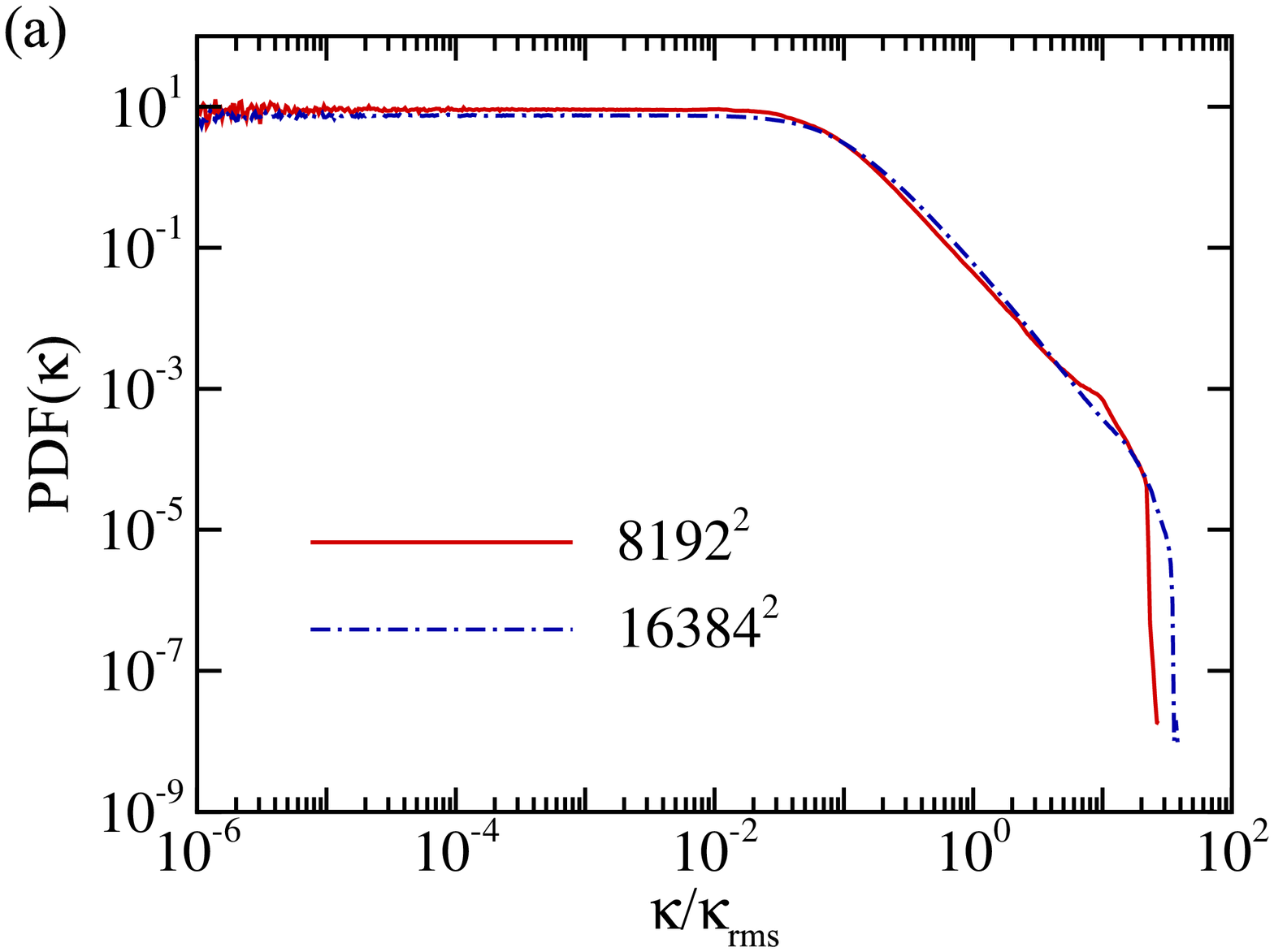}
\includegraphics[width=0.45\textwidth]{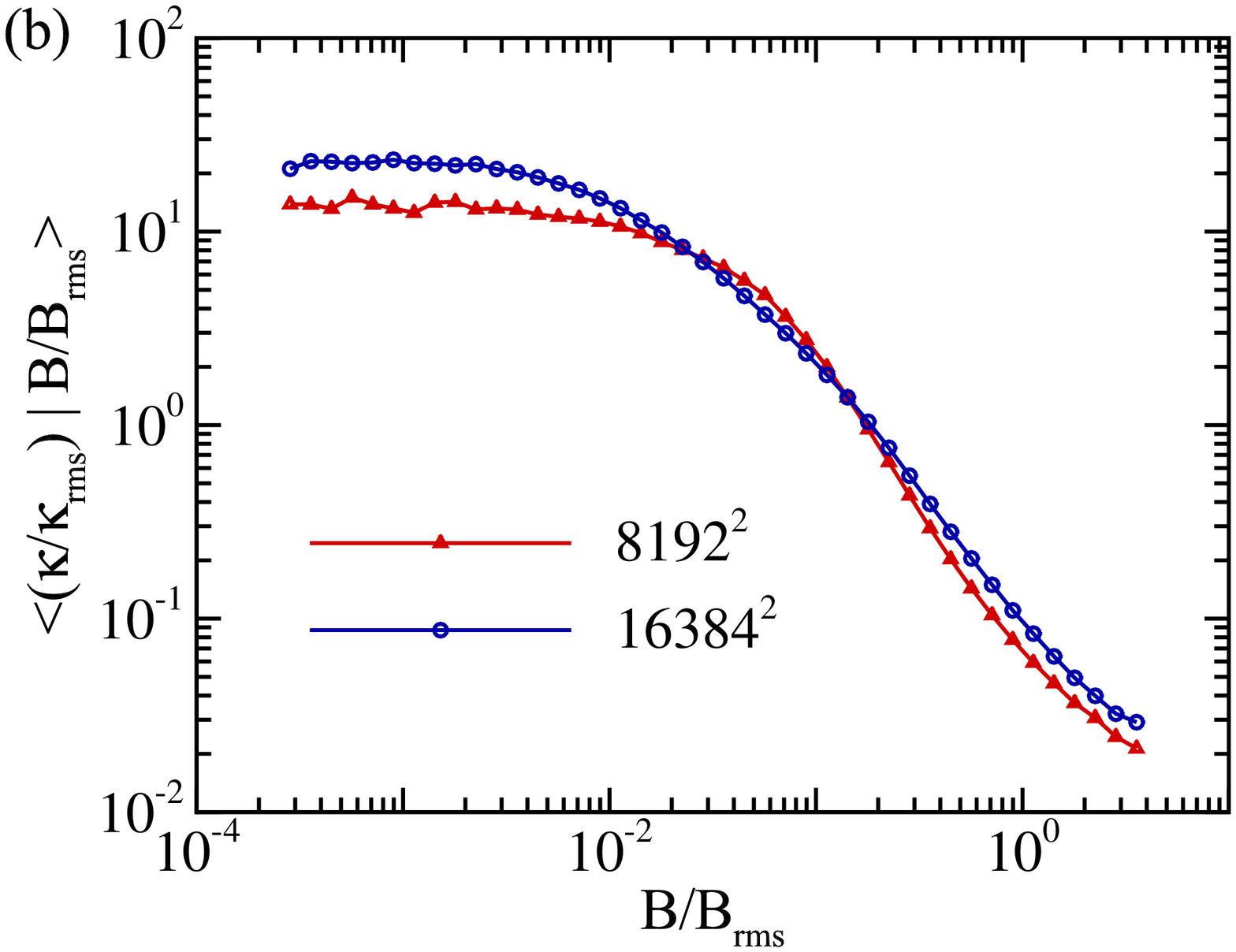}
\caption{Comparison between $8192^2$ resolution and $16384^2$ resolution:
(a) PDF of  magnetic field curvature $\kappa$;
(b) Average of magnetic field curvature $\kappa$ conditioned on magnetic magnitude.}
\label{fig:zero-padding}
\end{figure*}

The linear anti-correlation between curvature and the square of magnetic field
provides a plausible rationale for
the similarity between $\kappa B$ and $\kappa$ in Figs. \ref{xy-em-k}(b) and (d) ,
where here we also exclude from consideration
some special points like magnetic island cores and reconnection sites.
Although near both magnetic island cores and reconnection sites,
the direction of magnetic field
changes over very short length scales,
corresponding to intense curvature, the
product $\kappa B$ is not typically a local maxima at those
positions. This seems to be in apparent contradiction
to the preceding description of the general trend, that is,
the linear anti-correlation between curvature and the square of magnetic field
leads to $\kappa B$ in approximately linear relationship with $\sqrt{\kappa}$.
However, the measurement of curvature is inevitably limited by the numerical resolution of the simulation.
In order to illustrate the effect of grid resolution,
we apply a Fourier zero-padding and interpolation technique \cite{servidio2010statistics}
to obtain a $16384^2$ array in place of the original $8192^2$ array.
One can see from Fig. \ref{fig:zero-padding}(a) that
the Fourier zero-padding and interpolation technique improves the accuracy of
our measurements of curvature and higher resolution enables
resolution of more intense curvature.
Also noteworthy from Fig. \ref{fig:zero-padding}(b) is that the conditional
averages of $\kappa$ tend to form a plateau
as the magnetic field vanishes, which could be
near magnetic island cores and reconnection sites.
It is not possible to numerically measure true curvature therein
since there will always be noise as approaching these zero-magnetic positions.

\begin{figure*}%[!htpb]
\centering
\includegraphics[width=0.45\textwidth]{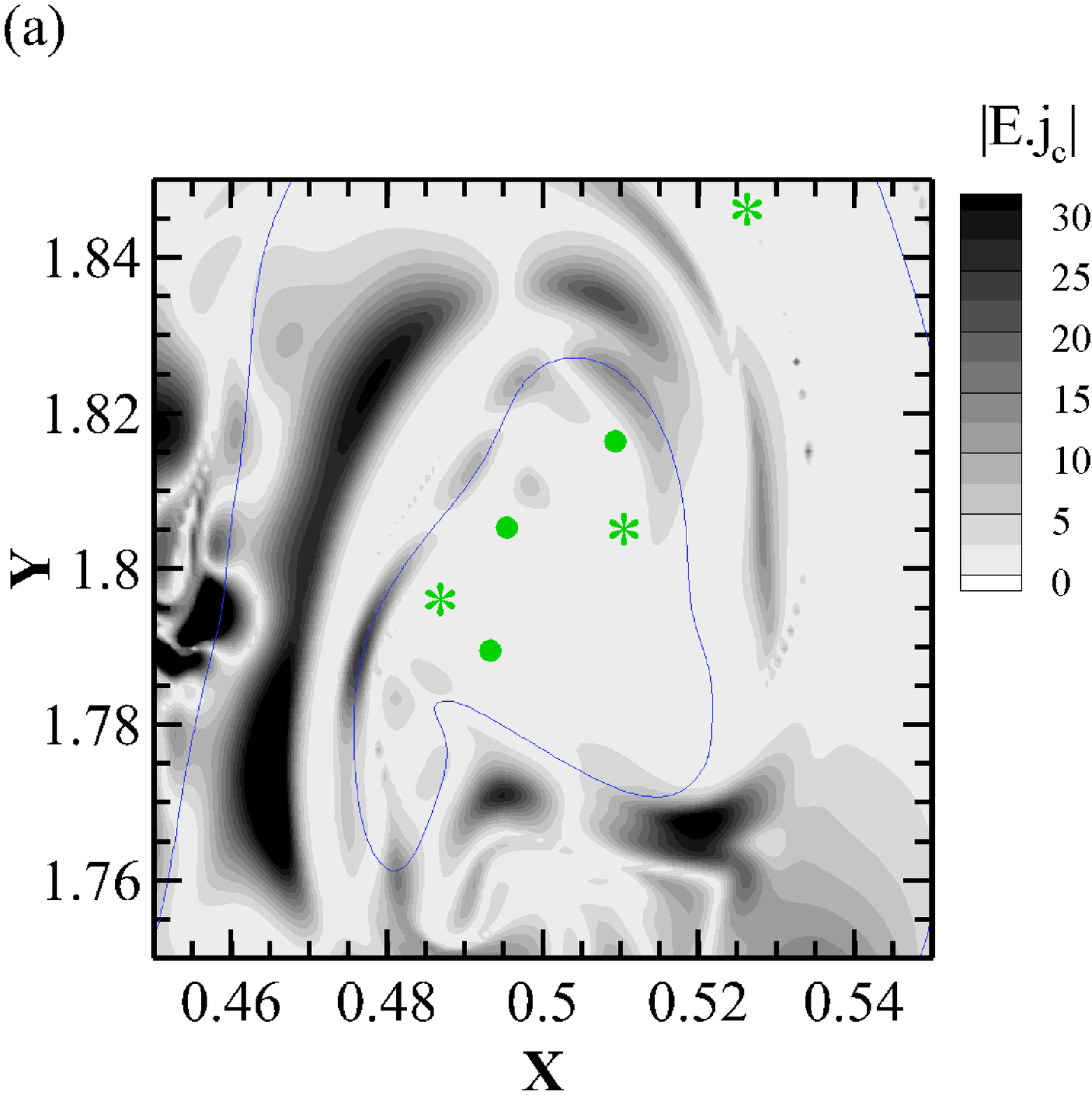}
\includegraphics[width=0.45\textwidth]{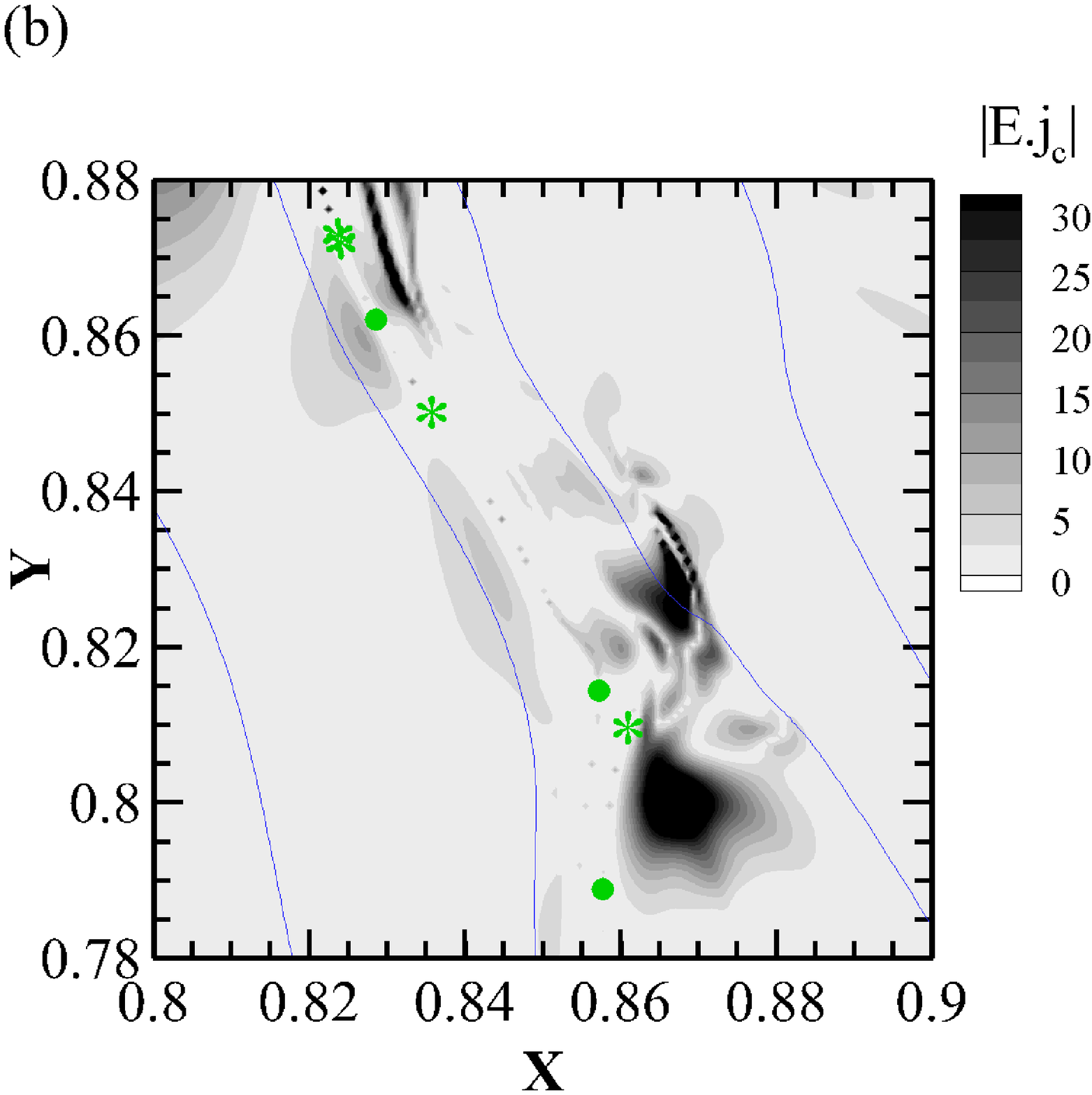}
\includegraphics[width=0.45\textwidth]{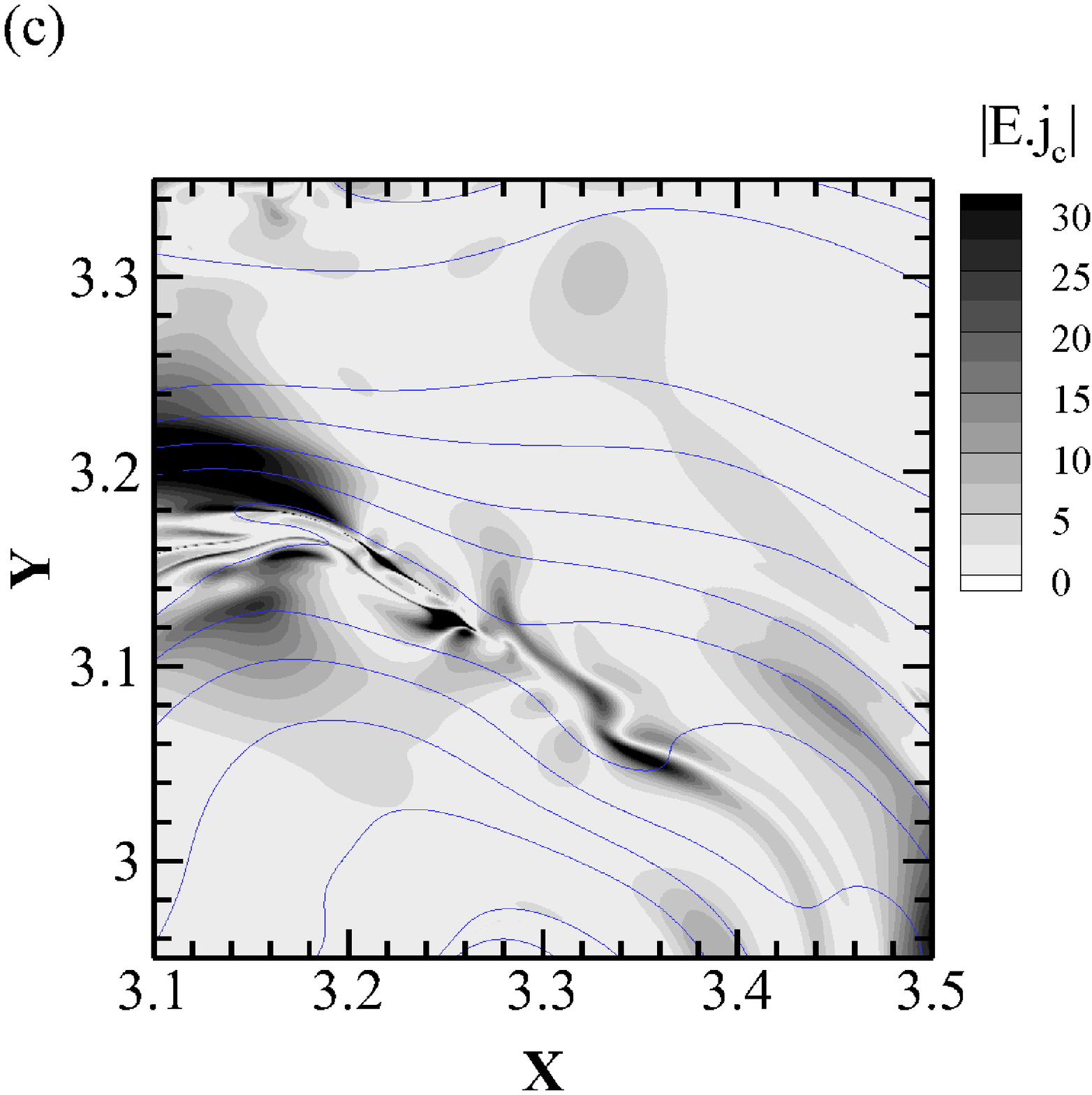}
\includegraphics[width=0.45\textwidth]{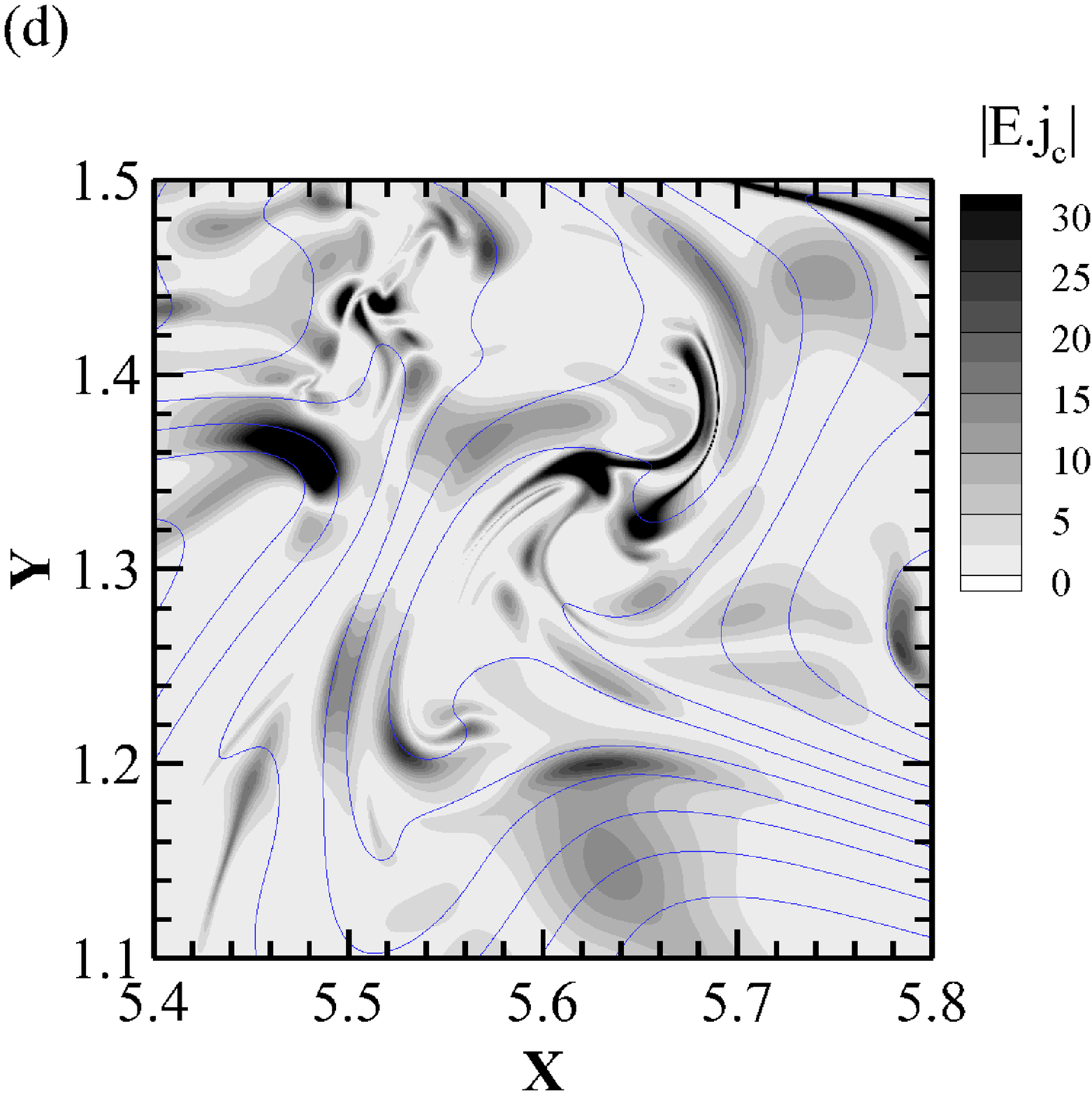}
\caption{Contour maps of the curvature drift acceleration term $|\boldsymbol{E} \cdot \boldsymbol{j}_c|$.
(a) and (b): regions with magnetic island cores marked as green circles
and magnetic reconnection sites marked as green stars.
(c) and (d): regions that do not contain topologically special points.
The blue contour lines
show the magnetic field lines.}
\label{xy-em-k-zoomin}
\end{figure*}

Previous studies \cite{dahlin2014mechanisms,dahlin2017role,li2017particle,li2015nonthermally,guo2014formation,lu2018formation,wang2016mechanisms,wang2017electron}
on magnetic reconnection
emphasize the prominence of curvature drift acceleration in reconnection exhausts,
at ends of contracting magnetic islands and in island merging regions.
The importance of this process also emerges by
virtue of the curvature analysis here.
Figs. \ref{xy-em-k-zoomin}(a) and (b) show the spatial distributions of
the curvature drift acceleration in several subregions, with
magnetic island cores marked as green circles and magnetic reconnection
sites marked as green stars.
On the one hand, the curvature drift acceleration at magnetic island cores and
reconnection sites is small, as we have shown, due to the product of
the perpendicular electric field $E_{\perp}$ and the curvature-related term $\kappa B$.
On the other hand, intense curvature is often found in the vicinity of these positions,
thus enhancing the curvature drift acceleration nearby.
The analyses we make so far make no specific reference to turbulence.
But turbulence and heating processes related to turbulence
are frequently implicated in the
study of space and astrophysical plasmas
\cite{Dmitruk04,TenBarge13,Perri12,price2016effects,Bandyopadhyay2018solar,yang2018scale}.
Evidently, turbulence, contemporaneous
with magnetic reconnection, operates cooperatively
in the natural evolution of the magnetic field in these
plasmas
\cite{Retino07,Sundkvist07,matthaeus2011needs,osman2014magnetic,zhdankin2013statistical,wan2014dissipation}.
Local flows in turbulence, even though they may not contain
a magnetic island core or a reconnection site, may also produce large curvature.
To isolate the effect of turbulence, we show regions away from topologically special points
in Figs. \ref{xy-em-k-zoomin}(c) and (d).
In comparison with the regions in the proximity of magnetic reconnection
sites and magnetic island cores,
the observable enhancement of the curvature drift acceleration emerges as well
in turbulence-dominant regions, supporting the view that turbulent effects
are playing an essential role in particle acceleration.

\section{Further applications}
In this section we discuss the properties of magnetic field curvature in two different systems
that deviate from the statistically homogeneous, isotropic and two-dimensional MHD:
isotropic 3D MHD and 2.5D kinetic plasma.

\subsection{3D MHD}
The incompressible three-dimensional MHD equations read
\begin{eqnarray}
{\frac{\partial \boldsymbol{v}}{\partial t}} + \boldsymbol{v} \cdot \nabla \boldsymbol{v} &=& -\nabla p^{*} + \boldsymbol{B} \cdot \nabla \boldsymbol{B} + \nu \nabla^2 \boldsymbol{v}, \\
{\frac{\partial \boldsymbol{B}}{\partial t}} + \boldsymbol{v} \cdot \nabla \boldsymbol{B} &=& \boldsymbol{B} \cdot \nabla \boldsymbol{v} + \eta \nabla^2 \boldsymbol{B},
\end{eqnarray}
where $p^*$ is the total (kinetic + magnetic) pressure, along with
$\nabla \cdot \boldsymbol{v}=\nabla \cdot \boldsymbol{B}=0$.
We solve the Fourier-space version of the above equations via
a Galerkin spectral method \cite{orszag1972numerical},
with $1024$ Fourier modes in each spatial direction.
For simplicity, equal viscosity and resistivity $\nu=\eta=4\times 10^{-4}$ are used.
The run is a freely decaying problem in periodic cube of size $2\pi$
and has the initially unity fluctuation energy equipartitioned between the kinetic and
magnetic components, i.e., $E_v=E_b=0.5$.
The fields are initialized at modes $1\le |\boldsymbol{k}| \le 5$
with random phases and fluctuation amplitudes, whose spectra are proportional to
$1/\left[1+(k/k_0)^{11/3}\right]$ with $k_0=3$. The cross-helicity is always small.
We carry out our analysis on a snapshot near the time of maximum mean square current density.

\begin{figure}
\centering
\includegraphics[width=0.5\textwidth]{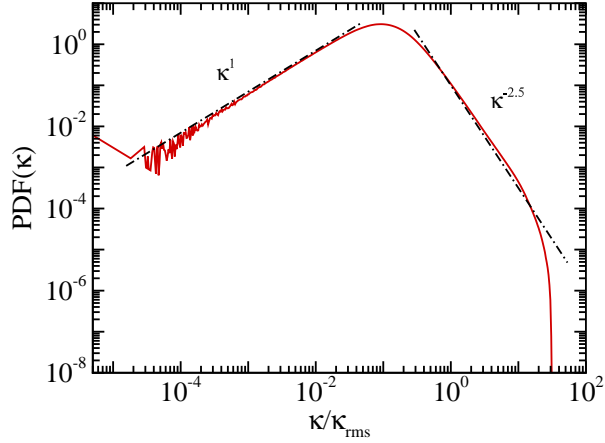}
\caption{3D MHD simulation: PDF of the magnetic field curvature $\kappa$ normalized to the root
mean square value $\kappa_{\rm rms}$. The PDF has a $\kappa^{1}$ low-curvature regime and a
$\kappa^{-2.5}$ high-curvature tail.}
\label{fig:3D-PDF-k}
\end{figure}

\begin{figure}
\centering
\centering
\includegraphics[width=0.45\textwidth]{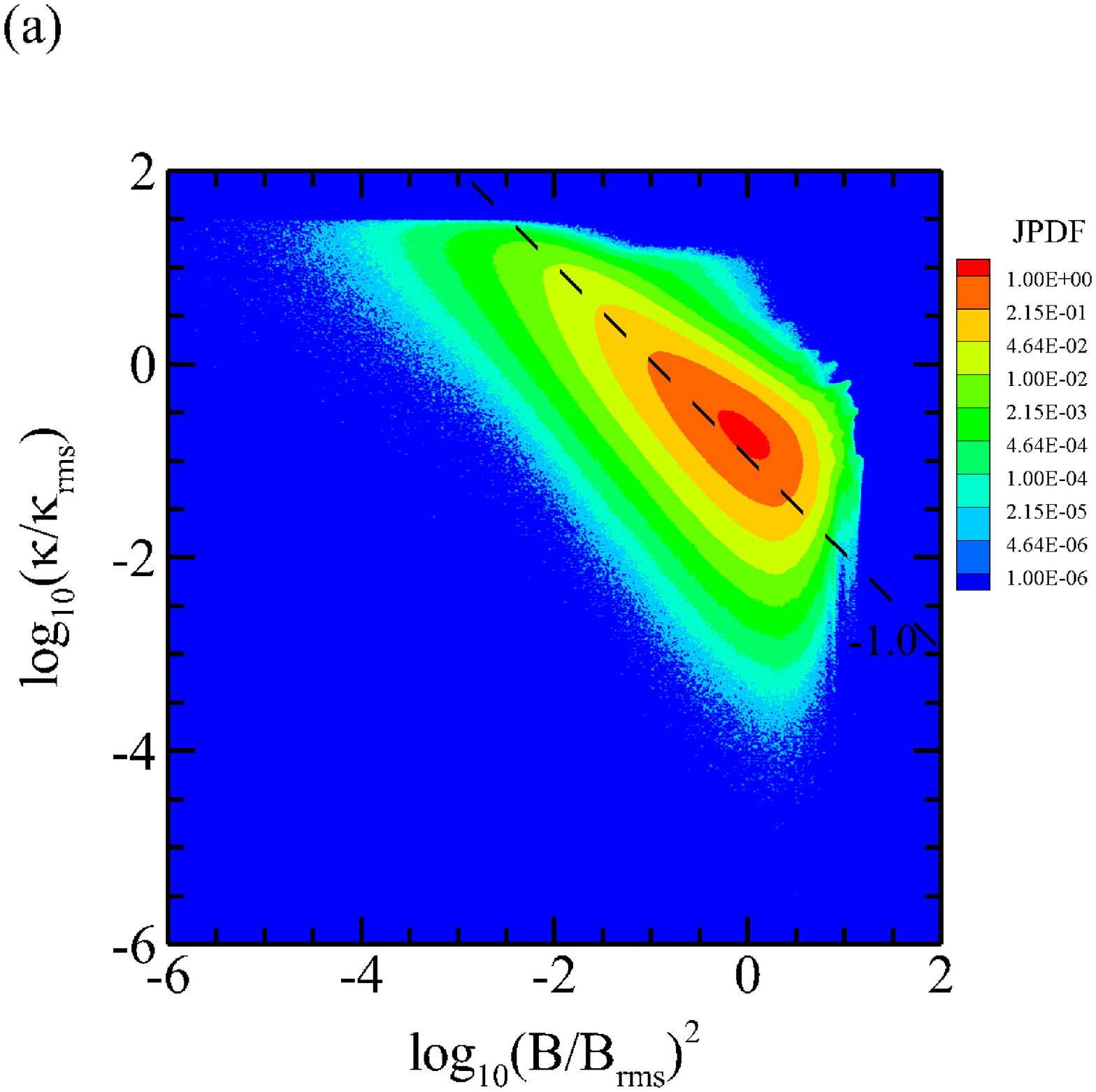}
\includegraphics[width=0.45\textwidth]{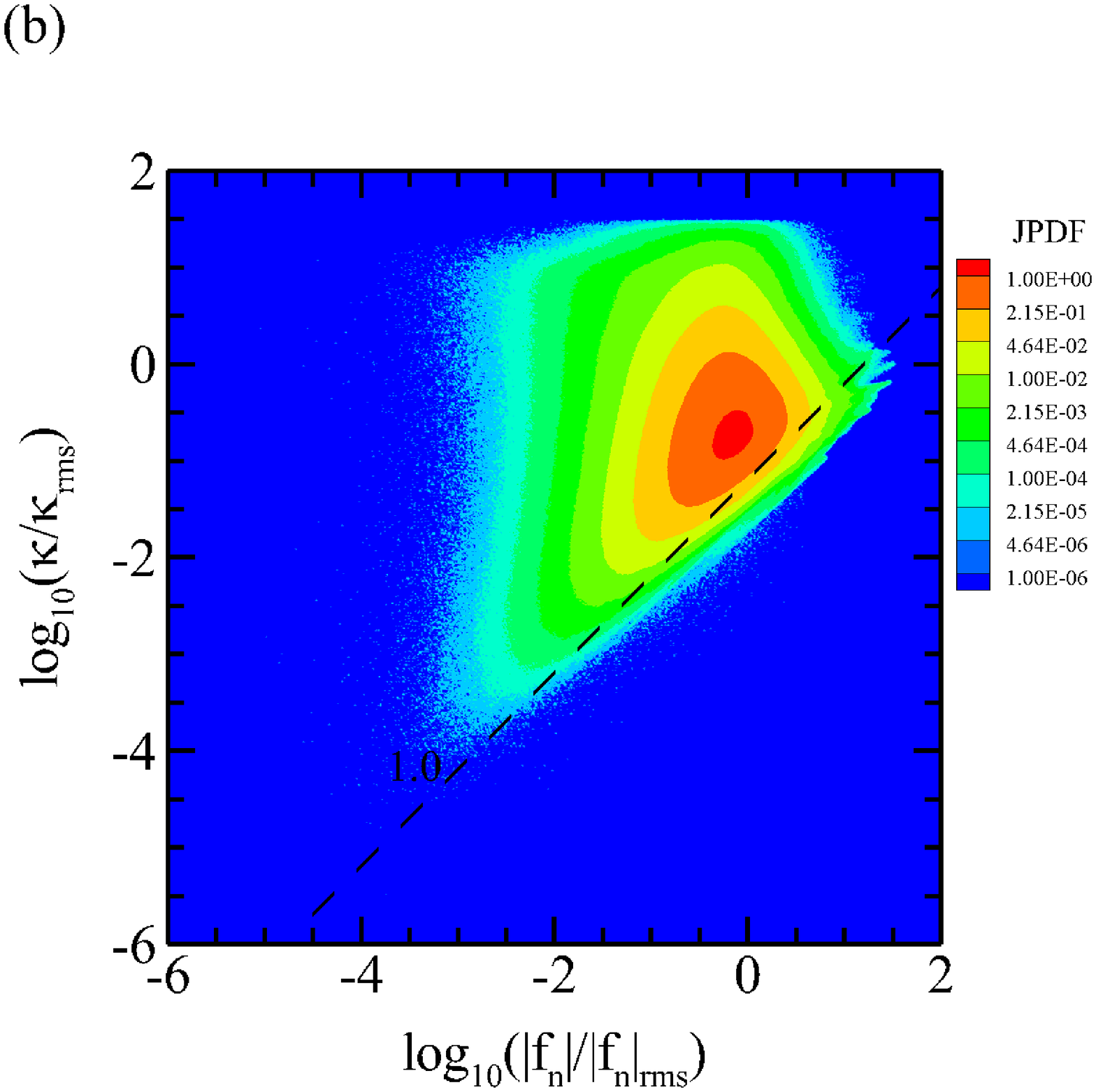}
\caption{3D MHD simulation: Joint PDFs of the curvature $\kappa$ and
(a) the square of magnetic magnitude $B^2$ and (b) the force magnitude $f_n$ acting normal to
field lines. All quantities are normalized to their respective root mean square values.
There are apparent associations between high curvature and low magnetic field
and between low curvature and low normal force.}
\label{fig:3D-JPDF-k-b}
\end{figure}

One can see from Fig. \ref{fig:3D-PDF-k} that the PDF of the curvature exhibits
power laws for both small-curvature and large-curvature regimes:
For small curvature, the PDF is close to linear with $\kappa$,
while, for large curvature, it scales as $\kappa^{-2.5}$.
Since dimensonality does not enter into
the expression in Eq. \ref{Eq.k},
we expect that the anti-correlation between curvature and magnetic field
will hold also in a 3D system. This expectation
is confirmed in our simulation, see Fig. \ref{fig:3D-JPDF-k-b},
which shows that low and high curvature is strongly correlated with
small normal force and small magnetic field magnitude, respectively.
In analogy with the procedure in 2D, the square of magnetic magnitude
$B^2$ in 3D MHD should then be distributed following a chi-squared distribution
with 3 degrees of freedom.
Since the curvature as $\kappa \rightarrow \infty$ scales like $B^{-2}$ as
$B^2 \rightarrow 0$, one can derive the scaling $P(\kappa) \sim \kappa^{-2.5}$ for high curvature.
Similar reasoning is applicable to the low-curvature regime.
Note that the force $f_n$ is confined to the plane orthogonal to the magnetic field.
We then assume the PDF of $f_n^2$ at small values is a chi-squared distribution with
2 degrees of freedom.
As $f_n \rightarrow 0$, $P(f_n) \sim f_n$, and we recover
$P(\kappa) \sim \kappa^{1}$ for low curvature.

\subsection{Kinetic plasma}
Plasma turbulence involves structures across a wide range of scales,
spanning from macroscopic fluid scales to sub-electron scales.
Based on what plasma properties we are interested in studying,
a plasma can be treated as tractable models in various limits.
MHD model remains a credible approximation for a plasma at scales large enough to be
well separated from kinetic effects, while more refined kinetic description is
required at kinetic scales.
Here we compare results from fully kinetic particle-in-cell (PIC) simulations
with those from MHD simulations.

We employ a fully kinetic simulation by P3D \cite{Zeiler02}
in 2.5D geometry (three components of dependent
field vectors and a two-dimensional spatial grid).
Number density is normalized to a reference
number density $n_r$ (=1 in this simulation),
mass to proton mass $m_i$ (=1 in this simulation),
charge to proton charge $q_i$, and
magnetic field to a reference $B_r$ (=1 in this run).
Length is normalized to the ion inertial length $d_i$,
time to the ion cyclotron time $\Omega_i^{-1}$,
velocity to the reference Alfv{\'e}n speed
$v_{Ar}=B_r/\left(4\pi m_i n_r\right)^{1/2}$, and
temperature to $T_r=m_i v_{Ar}^2$.
The simulation was performed in a periodic domain, whose size
is $L=150 d_i$, with $4096^2$ grid points and
$3200$ particles of each species
per cell ($\sim 107\times 10^9$ total particles).
The ion to electron mass ratio is $m_i/m_e = 25$,
and the speed of light in the simulation is $c=15 v_{Ar}$.
The run is
a decaying initial value problem,
starting with uniform density ($n_0=1.0$)
and temperature of ions and electrons ($T_0=0.3$).
The uniform magnetic field is $B_0 = 1.0$ directed
out of the plane.
More details about the simulation can be found in Ref. \onlinecite{yang2018scale}.
We analyze statistics using a snapshot near the time
of maximum root mean square electric current density.

\begin{figure}
\centering
\includegraphics[width=0.5\textwidth]{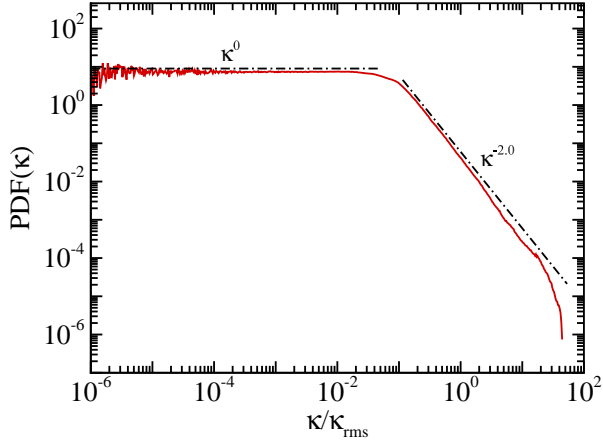}
\caption{PIC simulation: PDF of the magnetic field curvature $\kappa$ normalized to the root
mean square value $\kappa_{\rm rms}$. }
\label{fig:PIC-PDF-k}
\end{figure}

\begin{figure}
\centering
\centering
\includegraphics[width=0.45\textwidth]{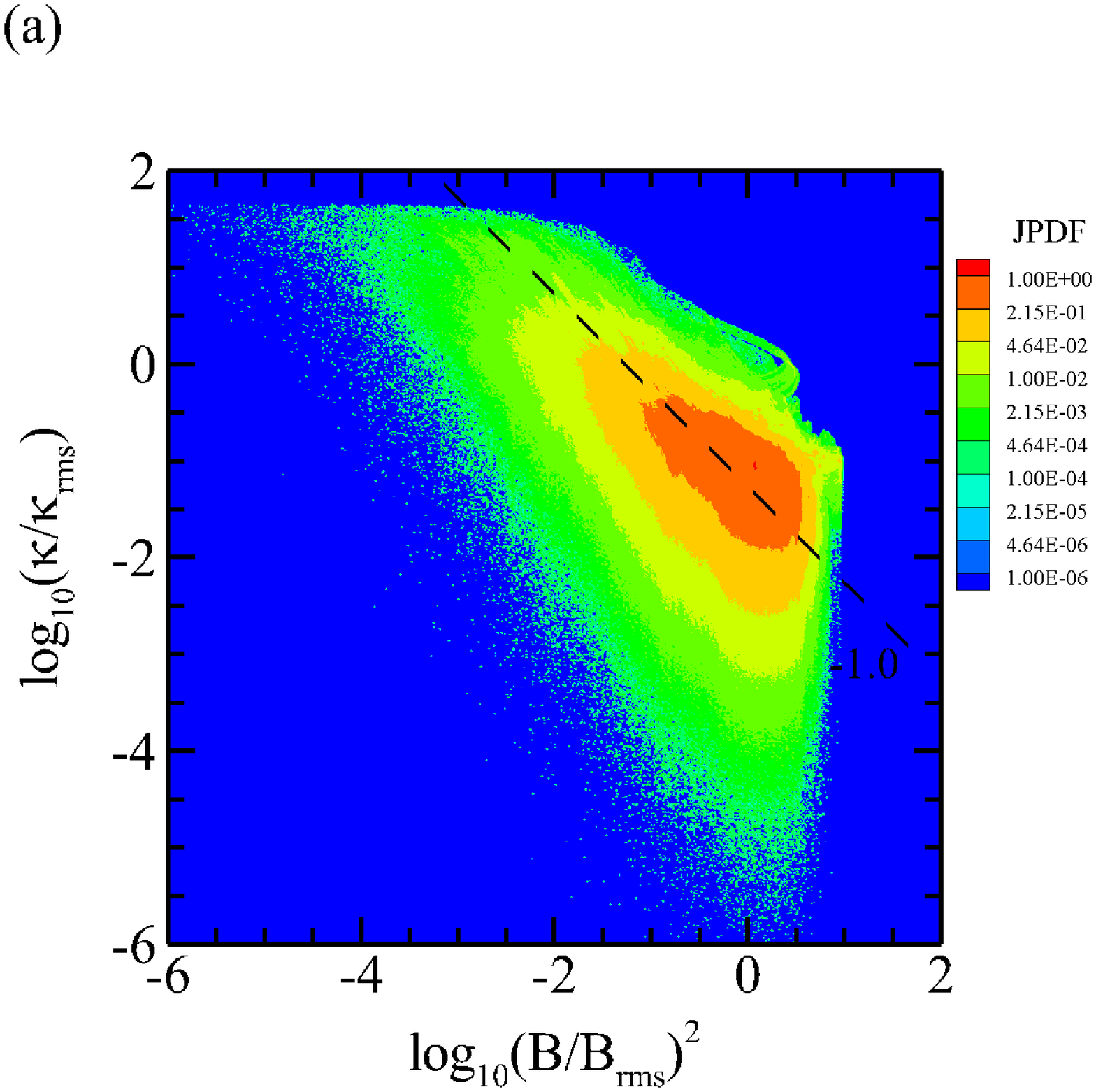}
\includegraphics[width=0.45\textwidth]{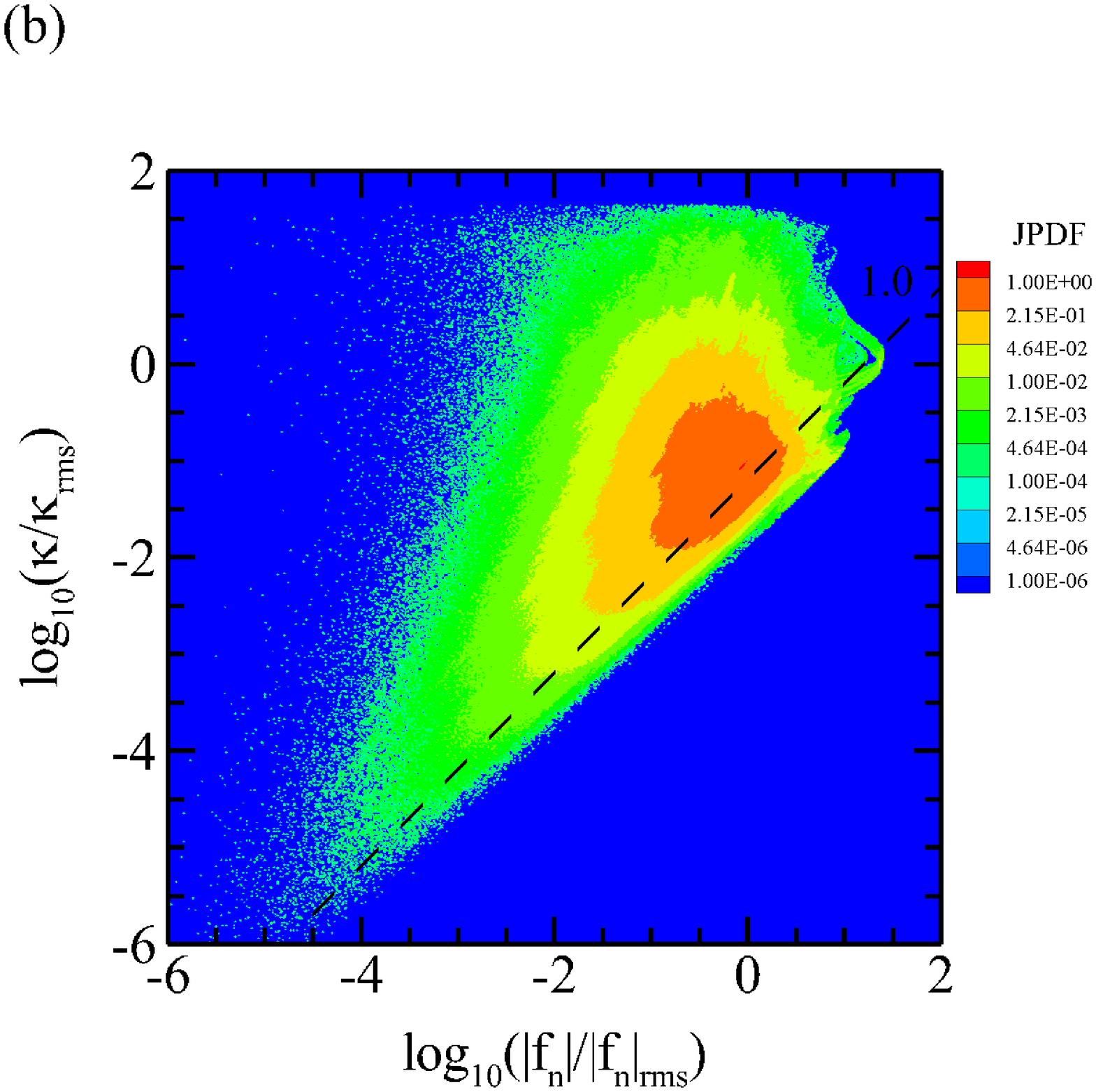}
\caption{PIC simulation: Joint PDFs of the curvature $\kappa$ and
(a) the square of magnetic magnitude $B^2$ and (b) the force magnitude $f_n$ acting normal to
field lines.}
\label{fig:PIC-JPDF-k-b}
\end{figure}

One can see from Fig. \ref{fig:PIC-PDF-k}  that
the PDF has a $\kappa^{0}$ low-curvature regime and a $\kappa^{-2}$ high-curvature tail.
Fig. \ref{fig:PIC-JPDF-k-b} indicates apparent associations
between high curvature and low magnetic field
and between low curvature and low normal force.
Although the PIC simulation enables the resolution of much smaller scales,
the corresponding behavior of curvature is essentially similar to that in 2D MHD.
The reasoning advanced in Sec. \ref{sec:2D-k-b} might therefore be deemed universal for
plasma turbulence.

\section{Conclusions}
Curvature characterizes magnetic field lines. For example,
both turbulence and magnetic reconnection drive tangled and bent
magnetic configurations, corresponding to intense curvature.
We can therefore analyze curvature properties to
improve understanding of the curvature drift mechanism,
often implicated in particle acceleration.
In this work, we have clarified
the dependence between high curvature and low magnetic field.
In particular, high curvature $\kappa$ is statistically scaled as $B^{-2}$, thus generating the power-law tails of the PDF of curvature. The curvature drift term, responsible for particle energization,
is found to be strongly associated with the
scalar curvature. It is active in high-curvature positions
which could be attributed to turbulence and magnetic reconnection.
We have not attempted to quantify the relative
strength of these two kinds of acceleration processes here.

The simulations used here include two-dimensional MHD and
three-dimensional MHD,
but it does not necessarily make a reference to the dimensionality when arriving at
Eqs. \ref{Eq.k} and \ref{Eq.curvature-drift},
in order to maintain as broad a context as possible.
It is therefore expected that the relevant statistical
features are substantially analogous in 3D and 2D.
Indeed, the anti-correlation between high curvature and low magnetic field applies as well
to the 3D case.
Also noteworthy is its application for kinetic plasma based on PIC simulation.
However, we should make it clear that this work is neither complete in its coverage nor exhaustive of possibilities.
The curvature drift mechanism we study is only
one possibility for particle acceleration
and other mechanisms have not been addressed herein.

%\begin{acknowledgments}
%We acknowledge
%\end{acknowledgment
\section{Acknowledgments}
This work has been supported by NSFC Grant Nos. 91752201 and 11672123;
the Shenzhen Science and Technology Innovation Committee (Grant No. JCYJ20170412151759222).
Yan Yang is supported by the Presidential Excellence Postdoctoral Fellowship from the Southern Univeristy
of Science and Technology.
WHM is supported
in part by NASA under the MMS Theory and Modeling team (NNX14AC39G),
and the Parker Solar Probe mission (Princeton subcontract SUB0000165).
We acknowledge computing support provided by Center for Computational 
Science and Engineering of Southern University of Science and Technology.

%\bibliography{Refs-MNRAS,Refs-Yan,Refs-whm-add}

\end{document}